\documentclass[11pt]{article}
\usepackage{authblk}
\usepackage[utf8]{inputenc}
\usepackage{aas_macros}
\usepackage{booktabs}
\usepackage{hyperref}
\usepackage{graphicx,color,rotating}
\usepackage{booktabs}
\usepackage{caption}
\usepackage{lineno,amssymb}
\usepackage[T1]{fontenc}

\title{{\bf Target of Opportunity Observations of Gravitational Wave Events with LSST}}
\author{The TVS Multiwavelength Characterization/GW Counterparts subgroup} 
\author{\\Raffaella Margutti (chair, Northwestern)} 
\author{\\Section leaders: P. Cowperthwaite (Carnegie)}
\author{Z.~Doctor (UChicago)}
\author{K.~Mortensen (Northwestern)}
\author{C.~P.~Pankow (Northwestern)}
\author{O.~Salafia (INAF)}
\author{V.~A.~Villar (Harvard)}
\author{\\Contributors (alphabetical order): K. Alexander (Northwestern)}
\author{J.~Annis (Fermilab)}
\author{I.~Andreoni (Caltech)}
\author{A.~Baldeschi (Northwestern)}
\author{B.~Balmaverde (UNIFI)}
\author{E.~Berger (Harvard)}
\author{M.~G.~Bernardini (INAF)}
\author{C.~P.~L.~Berry (Northwestern)}
\author{F.~Bianco (NYU)}
\author{P.~K.~Blanchard (Harvard)}  
\author{E.~Brocato (INAF)}
\author{M.~I.~Carnerero (INAF)}
\author{R.~Cartier (NOAO)} 
\author{S.~B.~Cenko (NASA/Goddard)}
\author{R.~Chornock (Ohio U)}
\author{L. Chomiuk (MSU)}
\author{C.~M.~Copperwheat (LJMU)}
\author{M.~W.~Coughlin (Caltech)}
\author{D.~L.~Coppejans (Northwestern)}
\author{A. Corsi (TTU)}
\author{F.~D'Ammando (INAF)}
\author{L.~Datrier (U Glasgow)}
\author{P.~D'Avanzo (INAF)}
\author{G.~Dimitriadis (UC Santa Cruz)}
\author{M.~R.~Drout (U Toronto/Carnegie)}
\author{R.~J.~Foley (UC Santa Cruz)}
\author{W.~Fong (Northwestern)}
\author{O.~Fox (STScl)}
\author{G.~Ghirlanda (INAF)}
\author{D.~Goldstein (Caltech)}
\author{J.~Grindlay (Harvard)}
\author{C.~Guidorzi (Ferrara)}
\author{Z.~Haiman (Columbia)}
\author{M.~Hendry (U Glasgow)}
\author{D.~Holz (UChicago)}
\author{T.~Hung (UC Santa Cruz)}
\author{C.~Inserra (Cardiff)} 
\author{D.~O.~Jones (UC Santa Cruz)}
\author{V.~Kalogera (Northwestern)}
\author{C.~D.~Kilpatrick (UC Santa Cruz)}
\author{G.~Lamb (Leicester)}
\author{T.~Laskar (U Bath)}
\author{A.~Levan (Warwick)}
\author{E.~Mason (INAF)}
\author{K.~Maguire (Belfast)}
\author{A.~Melandri (INAF)}
\author{D.~Milisavljevic (Purdue)} 
\author{A.~Miller (Northwestern)} 
\author{G.~Narayan (STScl)}
\author{E.~Nielsen (Stanford)}
\author{M.~Nicholl (U Edinburgh)}
\author{S.~Nissanke (UvA)}
\author{P.~Nugent (UC Berkeley)}
\author{Y.-C.~Pan (NAOJ)}
\author{D.~Pasham (MIT)}
\author{K.~Paterson (Northwestern)}
\author{S.~Piranomonte (INAF)}
\author{J.~Racusin (NASA/Goddard)}
\author{A.~Rest (STScl, JHU)}
\author{C.~Righi (Insubria)}
\author{D.~Sand (Arizona)}
\author{R.~Seaman (NOAO)}
\author{D.~Scolnic (U Chicago)}
\author{K.~Siellez (UC Santa Cruz)}
\author{L.~Singer (NASA/Goddard)}
\author{P.~Szkody (U Washington)}
\author{M.~Smith (Southampton)}
\author{D.~Steeghs (Warwick)} 
\author{M.~Sullivan (Southampton)}
\author{N.~Tanvir (Leicester)}
\author{G.~Terreran (Northwestern)}
\author{V.~Trimble (UC Irvine)}
\author{S.~Valenti (UC Davis)}
\author{with the support of the LSST Transient and Variable Stars Collaboration}
\affil{}

\date{November 30, 2018}

\begin{document}
\maketitle
\begin{abstract}
The discovery of the electromagnetic counterparts to the binary neutron star merger GW170817 has opened the era of GW+EM multi-messenger astronomy.
Exploiting this breakthrough requires increasing samples to explore the diversity of kilonova behaviour and provide more stringent constraints on the Hubble constant, and tests of fundamental physics.
LSST can play a key role in this field in the 2020s, when the gravitational wave detector network is expected to detect higher rates of merger events involving neutron stars ($\sim$10s per year) out to distances of several hundred Mpc. 
Here we propose comprehensive target-of-opportunity (ToOs) strategies for follow-up of gravitational-wave sources that will make LSST the premiere machine for discovery and early characterization for neutron star mergers and other gravitational-wave sources.
%
 \end{abstract}

\section{White Paper Information}
This white paper addresses:
\begin{enumerate} 
\item {\bf Science Category:} Exploring the transient and variable sky.
\item {\bf Survey Type Category:} Target of Opportunity observation.
\item {\bf Observing Strategy Category:} Integrated program with science that hinges on the combination of pointing and detailed observing strategy.
\end{enumerate}  

\clearpage
\section{Scientific Motivation}
\vskip -0.4 cm

The direct detection of gravitational waves (GW) from astrophysical sources has enabled an exciting new view of the cosmos \cite{GWdiscovery}. The true power of GW detections becomes apparent when they are paired with  electromagnetic (EM) data.
The identification of an EM counterpart provides numerous benefits including: improved localization leading to host-galaxy identification, determination of the source's distance and energy scales, characterization of the progenitor local environment, breaking modeling degeneracies between distance and inclination, and insight into the hydrodynamics of the merger. Furthermore, identification of the EM counterpart facilitates other fields of study such as determining the primary sites of heavy r-process element production, placing limits on the neutron star (NS) equation of state, and making independent measurements of the local Hubble constant \cite{AbbottNSdiscovery,AbbottH0,Hjorth17}.

Exploiting the success of multi-messenger astronomy in the next decade will require the continued investment of observational resources. In this period, Advanced LIGO and Virgo (ALV) will increase their sensitivity, while additional interferometers will come online (KAGRA and LIGO-India) \cite{2018LRR....21....3A}. In this multi-detector regime, binary neutron star (NS-NS) mergers will be detected to $\sim$$200$~Mpc (and several hundred Mpc for neutron star -- black hole (NS-BH) binaries) and typical localizations will continue to improve from $\sim$$100$~deg$^2$ to $\sim$$10$~deg$^2$.
In this regime, LSST will provide a unique combination of large aperture and wide field-of-view that will be well suited to the task of GW follow-up. LSST will be able to cover typical GW localization regions in just a handful of pointings and achieve deep observations with relatively short integration times. This means that LSST will be able to detect and identify EM counterparts to GW sources more effectively and rapidly,
particularly those at the 
largest distances.

In this white paper, we propose comprehensive target-of-opportunity strategies for the follow-up of GW sources that will allow LSST to serve as the premiere discovery instrument in the Southern Hemisphere. We outline two strategies: a {\it minimal} strategy that targets a time investment of $\sim 1\%$ of the total survey time ($\sim 40$~hr~yr$^{-1}$) and an {\it ideal} strategy that will use $\sim 2\%$ of the total survey time ($\sim 80$~hr~yr$^{-1}$). These strategies are designed to tackle the following major science goals:

\textbf{[i]} The primary science goal for studies of EM transients from GW sources in the 2020s will be growing the sample size of known events, with a strong focus on finding the faintest events at the edge of the ALV detection horizon. Targeted follow-up will be much more efficient at achieving this goal compared to waiting for serendipitous discoveries from the LSST Wide-Fast-Deep (WFD) survey (e.g. \cite{Cowperthwaite18}). Building  a large sample of EM counterparts is essential for conducting statistically rigorous systematic studies that will allow us to understand the diversity of EM behaviors, their host environments, the nature of merger remnants, and their contribution to the chemical enrichment of the universe through cosmic $r$-process production, which shapes the light-curves and colors of ``kilonovae'' (KNe) associated to GW events (e.g., \cite{2015MNRAS.446.1115M}).

\textbf{[ii]} Of particular interest are observations of KNe at very early times (e.g., $\lesssim10$~hr post-merger). Despite the fact that the optical counterpart of GW170817 was discovered less than 11 hours post-merger (e.g., \cite{Arcavi17,Coulter17,Cowperthwaite17,Drout17,Kasliwal17,Lipunov17,Smartt17,Soares17,Tanvir17,Valenti17,Villar17}), these observations were still unable to definitively determine the nature of the early time emission. Understanding this early-time emission is crucial for identifying emission mechanisms beyond the KN (e.g., neutron precursor, shock-cooling, e.g., \cite{PiroKollmeier18}). In particular, mapping the rapid broad-band SED evolution
is key to separating these components, and also distinguishing
KN from most other astrophysical transients.

\textbf{[iii]} Yet to be observed is the merger of a NS-BH. This system
is expected to produce a KN, but depending on the 
mass ratio of the binary and the NS equation of state
 there may be less or more material ejected (e.g. \cite{Foucart2018})
It is also unclear if NS-BH mergers will be able to produce the bright early-time blue emission seen in GW170817 \cite{2015MNRAS.446.1115M}. Furthermore, these systems will be gravitationally louder and thus GW-detected on average at greater distances. This combination of increased luminosity distance and potentially fainter counterpart means that LSST will be an essential tool for discovering their EM counterparts.

\textbf{[iv]} LSST has the potential to place deep limits on the optical emission from binary black holes (BH-BH) mergers. There are numerous speculative mechanisms for the production of an optical counterpart to a BH-BH merger, \cite{Perna,Loeb,Stone,deMink,McKernan}
yet none has been unambiguously observed. LSST will be able place deep limits on the optical emission from BH-BH mergers with a high statistical confidence in the case of non-detections, or might be able to discover the first EM counterpart to BH-BH mergers.

\textbf{[v]} Lastly, LSST has the capabilities to explore the currently uncharted territory of EM counterparts of unidentified GW sources.

\vskip +0.1 cm

In the pursuit of these goals, the true power of LSST will be the ability to both rapidly grow the population of known transients (e.g., KNe) and discover new sources of optical emission associated with compact object mergers (e.g., non-KN early-time emission, emission from a BH-BH merger) and unidentified GW sources.
\clearpage
\section{Technical Description}
\begin{footnotesize}
{\it Describe your proposed observations. Please comment on each observing constraint
below, including the technical motivation behind any constraints. Where relevant, indicate
if the constraint applies to all requested observations or a specific subset. Please note which 
constraints are not relevant or important for your science goals.}
\end{footnotesize}

\subsection{High-level description}
\begin{footnotesize}
\vskip -0.3 cm
{\it Describe or illustrate your ideal sequence of observations.}
\end{footnotesize}

In the ``LSST Observing Strategy'' document publicly released in August 2017 (which preceded the electromagnetic detection of GW170817 by just a few weeks), we conducted a preliminary model-agnostic study to assess the potential role of the LSST WFD survey in the follow-up of GW events (Chapter 6.5). Using the current cadence \texttt{minion1016} we quantified the probability that baseline LSST operations within the WFD survey would observe a GW localization region by pure chance multiple times within $\sim1$ week since the GW trigger, finding that the likelihood is extremely small ($\sim7\%$ for  $r$-band only; $\sim$ a few $\%$ for observations in multiple filters). We concluded that LSST can be the premier player in the southern hemisphere in the current era of multi-messenger astrophysics with GWs \emph{only if} equipped with ToO capabilities. 

 These conclusions have been significantly strengthened by recent detailed studies published in \cite{Cowperthwaite18,scolnic2017}, which focused on the problem of the detection and characterization of KNe from NS-NS mergers in the LSST WFD data stream using realistic simulations of the observing cadence and conditions. These studies either started from re-scaled versions of  the single known KN event with multi-band light-curves \cite{scolnic2017}, or simulated the expected KN light-curves for a wide range of ejecta properties and composition \cite{Cowperthwaite18}.  The main findings from these studies can be summarized as follows: (i) The main LSST survey will reach an overall efficiency of KN detection of the order of $\sim$ a few $\%$.\footnote{The definition of what constitutes a detection varies from study to study, but generically implies the capability to detect with high statistical confidence the KN emission in one or multiple bands and in at least one instance in time, and reject asteroids.} For a NS-NS merger rate $R \sim 1500$\,Gpc$^{-3}$\,yr$^{-1}$ \cite{NSNSrate} this result implies that  $< 7$ KNe  per year will be detected in the LSST WFD, and $\sim0.5$ KNe per year in the LSST Deep-Drilling Fields (DDFs) \cite{scolnic2017,Cowperthwaite18}. (ii) While the number of $\le7$ detections per year might seem encouraging, in reality the vast majority of the detected KNe will have poorly sampled light-curves, which will prevent accurate
 estimates of the merger ejecta, a physical parameter of  primary scientific importance.

 These two results described in (i)+(ii) are direct consequences of the fact that the cadence of the LSST WFD survey is a poor match to the fast time-scales of evolution of GW counterparts,
 and that the sky area covered by the DDFs is not large enough to rely on chance alignment with GW localizations. ToO capabilities are the only way to enable LSST to have a significant scientific role in the new era of Multi-Messenger Astrophysics with GWs.

Below we analyze separately the cases of ToO follow-up of GW triggers resulting from NS-NS mergers, NS-BH mergers and binary black hole (BH-BH) mergers. For each of these classes we outline a ``minimal'' and  ``optimal'' LSST follow-up strategy. We design the LSST follow-up strategy of GW triggers bearing in mind that at the time of writing we have only one example of well observed KN from the NS-NS merger event GW170817 (unambiguous EM counterparts to NS-BH and BH-BH mergers are yet to be found), and that our knowledge of EM counterparts to GW events is likely to dramatically improve in the next few years \emph{before} the start of LSST operations. 

In the 2022+ era of LSST operations, the sky localization regions from a three GW-detector network operating at design sensitivity will routinely be of the order of  $20$--$200$ deg$^2$, depending on distance, sky location and orientation of the merger event \cite{2018LRR....21....3A}. The addition of the Japanese detector KAGRA at design sensitivity
and possibly LIGO India will further shrink the area and although their impact (and timeline) are uncertain, areas of 10s of deg$^2$ may be common.
LSST has a unique combination of capabilities for optical/NIR counterpart searches: the $\sim$10 deg$^2$ camera, deep sensitivity (over 6 bands) and a deep sky template for subtraction after the first year of operations. 
In addition, the incredibly fast readout and slew times are ideally suited to fast mapping of the expected $20$--$200$ deg$^{2}$ areas to depths that are untouchable by the other surveys currently in this search and discovery mission. Facilities such as ATLAS, ZTF and GOTO can cover large areas with their cameras, but do not have the aperture to go beyond magnitude $21$--$22$ and have limited filter sets. Pan-STARRS (in the northern hemisphere) and DECam (in the southern hemisphere) are more sensitive. 
Compared to DECam, LSST has the following key advantages: a larger FOV, larger collecting area, which makes LSST significantly more sensitive, shorter read-out time and the advantage of having an all-sky reference frame with which to do immediate image subtraction. The same arguments of LSST having the more powerful combination of sensitivity at a large FOV apply to other planned facilities such as BlackGEM and a southern hemisphere GOTO node.

\subsubsection{Binary Neutron Star mergers (NS-NS)}
\label{SubSubSec:NSNS}
For NS-NS mergers we identify two key areas of the parameter space that can \emph{only} be explored by LSST: (i) the very early ($\delta t<12$ hrs) multi-band evolution of the KN emission; (ii) the faint end of the KN brightness distribution (either resulting from distant NS-NS mergers detected by the GW interferometers, or from intrinsically low-luminosity events that populate the faint end of the KN luminosity function). We design the LSST follow-up strategy of NS-NS mergers around the two discovery areas above.

By sampling the rise time of the KN emission in multiple bands, LSST will enable constraints on new emission components
in KNe, like shock cooling emission (proposed for GW170817 by \cite{PiroKollmeier18}) or a neutron star precursor \cite{2015MNRAS.446.1115M}. 
The LIGO Scientific \& Virgo Collaboration (LVC) recently updated the predictions for sky areas and rates of detections in \cite{2018LRR....21....3A}. LSST is expected to start operations in 2022, before the earliest possible date that LIGO India may be operational ($\sim2024$ according to \cite{2018LRR....21....3A}).  Considering the three detectors of the LVC, the median  localization sky area at the start of the LSST operational years in 2022 is $110$--$180$ deg$^{2}$, with $14$--$22\%$ of merger events having a 90\% confidence region $\Omega_{\rm{90\%}}<$20 deg$^{2}$. This is a conservative estimate as it does not fold in the benefits of KAGRA joining the network. The range distance for NS-NS mergers is expected to be $\sim$ 200 Mpc. With the addition of KAGRA (expected to be at design sensitivity by 2024) and LIGO India 
the median $\Omega_{\rm{90\%}}$ significantly improves to $\Omega_{\rm{90\%}}=9$--$12$ deg$^{2}$ and 65-73\% of NS-NS mergers will have $\Omega_{\rm{90\%}}<$20 deg$^{2}$. 

In this document we design our strategies based on the expected performance of the GW detectors in 2022 according to \cite{2018LRR....21....3A}, keeping in mind that GW localizations might significantly improve as early as in 2024.
LSST can thus image the entire NS-NS localization region with a relatively small number of pointings (overlapping pointings to cover chip gaps will be necessary). This implies that LSST will be able to capture the multi-band evolution of KNe potentially starting as early as minutes after the GW trigger.
 Other survey instruments do not reach a comparable depth and, because of their smaller FOV, will have to tile the GW localization region with several pointings. Both factors will effectively prevent the systematic exploration of the very early KN evolution by ground-based instruments other than LSST. The multi-band exploration of the very early KN emission is a key strength of the LSST GW follow-up program that we propose here.

\begin{figure}
\center{\includegraphics[scale=0.3]{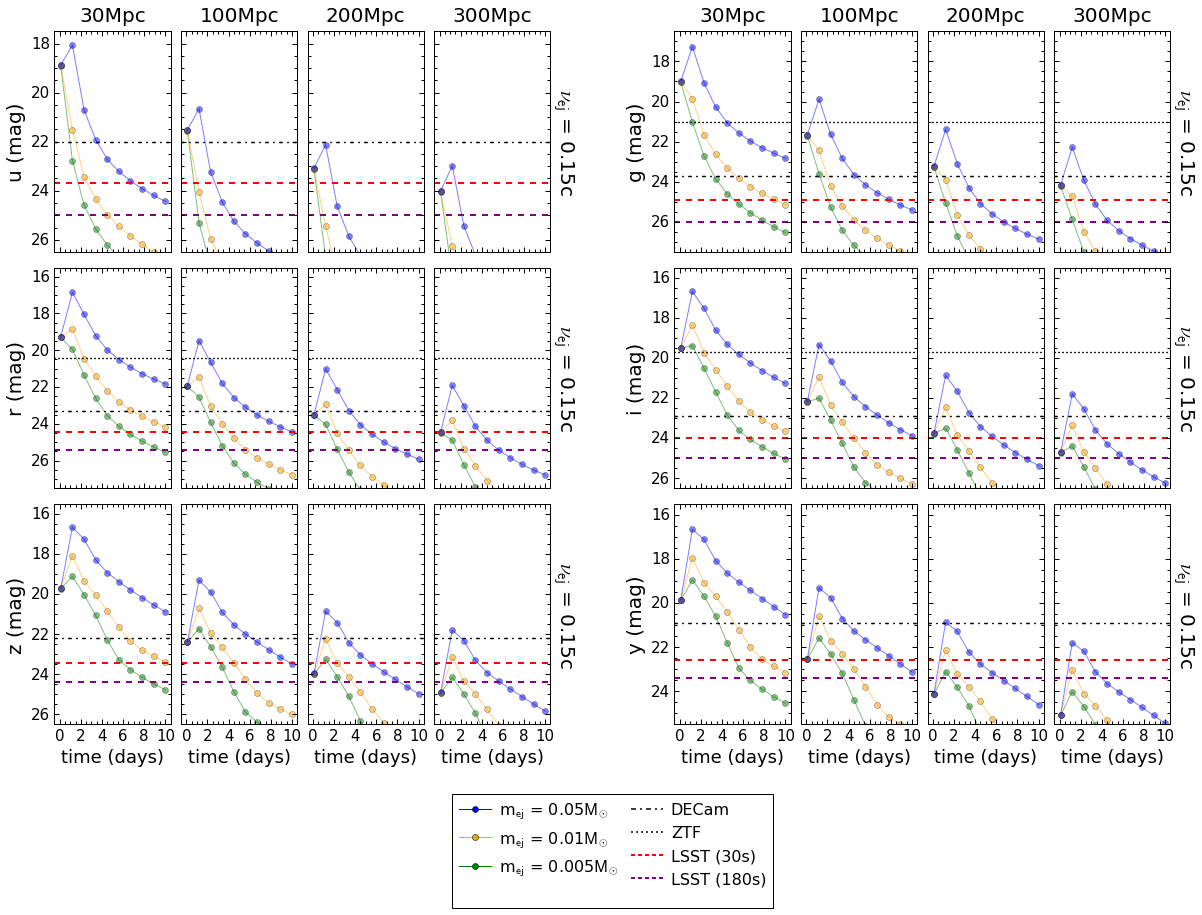}}
\caption{Simulated kilonova (KN) light-curves in the six LSST filters for different properties of the ejecta (mass and velocity) at four representative distances (30, 100, 200 and 300 Mpc). The models include a ``red''  and ``blue'' KN component. We explore three values of the red KN ejecta  mass $M_\mathrm{ej,R}=0.005,0.01, 0.05\,\rm{M_{\odot}}$ and velocity $v_\mathrm{ej,R}=0.15\,c$ (the KN luminosity is not a strong function of $v_\mathrm{ej,R}$ and values within $0.1$--$0.2\,c$ give comparable results). For each combination of these parameters the blue ejecta component is $M_\mathrm{ej,B}= 0.5
\times M_\mathrm{ej,R}$ and $v_\mathrm{ej,B}= 1.5\times v_\mathrm{ej,R}$. Dotted and dot-dashed horizontal lines mark the $5\sigma$ threshold of detection of ZTF and DECam, respectively. Red and purple dashed lines: $5\sigma$ LSST threshold of detection for exposure times of 30 s and 180 s under ideal observing conditions.  Adapted from Mortensen et al., in prep., to include the results from \cite{Cowperthwaite18}.}
 \label{Fig:KNredblue}
\end{figure}

\begin{figure}
\center{\includegraphics[scale=0.3]{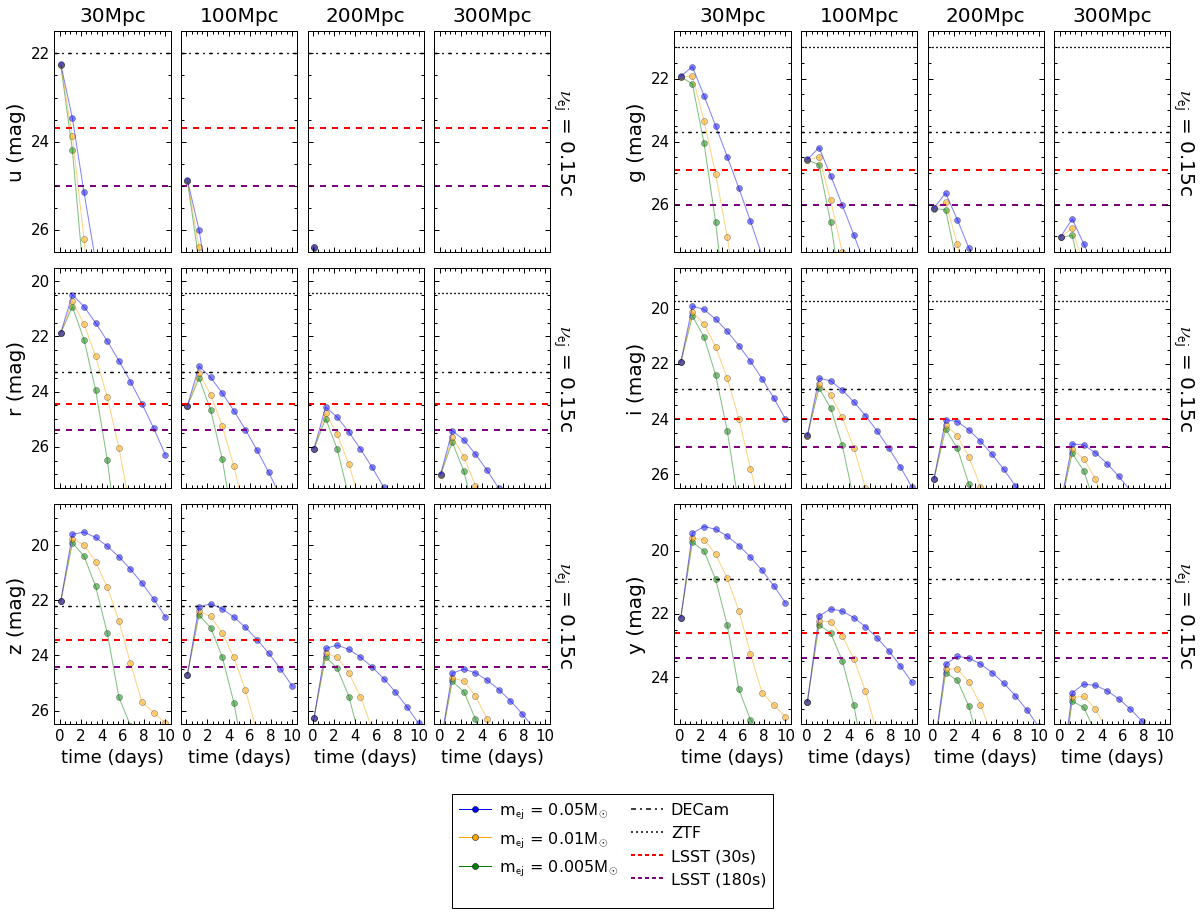}}
\caption{Simulated red kilonova (KN) light-curves in the six LSST filters for red ejecta mass $M_{\rm{ej,R}}=0.005,0.01, 0.05\,\rm{M_{\odot}}$ and velocity $v_{\rm{ej,R}}=0.15\,c$  at four representative distances (30, 100, 200 and 300 Mpc).  Dotted and dot-dashed horizontal lines mark the $5\sigma$ threshold of detection of ZTF and DECam, respectively. Red and purple dashed lines: $5\sigma$ LSST threshold of detection for exposure times of 30 s and 180 s under ideal observing conditions.  Adapted from Mortensen et al., in prep., to include the results from \cite{Cowperthwaite18}.}
 \label{Fig:KNred}
\end{figure}

A second key strength of our proposal builds on the unique LSST capability to map the faint end of the KN brightness distribution. Faint KN emission might result from the most distant NS-NS mergers detected by the GW interferometers, or from intrinsically low-luminosity nearby events (e.g. events with significantly smaller ejecta masses than GW170817). In the LSST era, NS-NS mergers are expected to be detected out to $\sim200$ Mpc. As shown in Fig. \ref{Fig:KNredblue} and \ref{Fig:KNred}, at this distance LSST is the only survey instrument able to detect red KNe and map their evolution for two weeks after GW trigger, enabling an accurate estimate of the ejecta parameters (mass, velocity).
This is a crucial aspect of EM follow-up of NS-NS mergers as: (i) the blue KN component is not guaranteed to be present in all NS-NS mergers \cite{2015MNRAS.446.1115M}; (ii) even if present, the brightness of the blue KN component is angle-dependent, and will thus depend on our line of sight to the NS-NS merger \cite{Kasen17}.  A solid discovery strategy of EM counterparts to NS-NS mergers has thus to be built around the capability to detect the red KN component.  As shown in Fig.~\ref{Fig:KNred}, the red emission from KNe at $200$ Mpc and with small ejecta mass  $M_\mathrm{ej,red}=0.005\,\rm{M_{\odot}}$ ($\sim$ half the ejecta mass inferred for the red KN associated with GW170817, e.g.\ \cite{Cowperthwaite17,Drout17,Smartt17,Villar17}) is well within the reach of one LSST visit, while it is beyond or at the very limit of what other instrument surveys will be able to detect. LSST observations of KNe will allow us to probe the \emph{diversity} of the ejecta properties of NS-NS mergers in ways that are simply not accessible otherwise.   

Below we outline our minimal and optimal LSST observing strategies of NS-NS mergers adopting a NS-NS merger rate $R\sim 300-5000\,\rm{Gpc^{-3}yr^{-1}}$, which translates into an expected number of NS-NS detections of $4-80\,\rm{yr^{-1}}$ with the three GW-detector network \cite{NSNSrate}.

\textbf{MINIMAL STRATEGY:} We propose multiple $u+g+r+i+z+y$ visits (30 s for each filter) of well-localized NS-NS mergers with $\Omega_{\rm{90\%}}\le20\,\rm{deg^2}$ and for which the sky position and time are favorable for prompt follow up (i.e. within hours since GW trigger) and continued follow up during the first night. Observations will be log-spaced in time with focus on the first
night the object is available to sample the very early KN evolution. The  $u$-band is of particular interest as there are predictions of a free-neutron decay pulse within the first few hours after merger \cite{2015MNRAS.446.1115M}. We will aim at 1 hr, 2 hr and 4 hr and, if the field is optimally placed, at 8hrs since GW trigger, followed by observations at 1 and 2 days. Based on the results from \cite{Cowperthwaite18}, we further propose $g+z$ observations of NS-NS mergers localized within $20\,\rm{deg^2}<\Omega_{\rm{90\%}}\le100\,\rm{deg^2}$ (30 s for each filter) with the same cadence as above. The $g+z$ filters are chosen to sample the widest possible range of the EM spectrum while maximizing the sensitivity of the observing campaign of less well localized targets (e.g. avoiding the throughput losses of the $u$ and $y$ filters). We expect to have identified the EM counterpart by the first or second night, at which point a public announcement would be immediately made, allowing all large area (but smaller FOV) facilities (e.g., VLT, Keck, Gemini, Magellan etc) to take over. 

On average we anticipate that $N=4$ ($N=20$) LSST pointings will be needed to cover the localization area of mergers with $\Omega_{\rm{90\%}}\le20\,\rm{deg^2}$ ($\Omega_{\rm{90\%}}\le100\,\rm{deg^2}$). With this strategy we expect to spend on average $\sim2.17$\,hrs and $\sim2.28$\,hrs per NS-NS merger with $\Omega_{\rm{90\%}}\le20\,\rm{deg^2}$ and $20\,\rm{deg^2}<\Omega_{\rm{90\%}}\le100\,\rm{deg^2}$, respectively.  Based on the results presented in \cite{2018LRR....21....3A}, the number of LSST accessible mergers with $\Omega_{\rm{90\%}}\le20\,\rm{deg^2}$ is $N\le (1-2)\,\rm{yr^{-1}}$,  while for $20\,\rm{deg^2}<\Omega_{\rm{90\%}}\le100\,\rm{deg^2}$ $N=(1-10)\,\rm{yr^{-1}}$, leading to a yearly LSST average time request for NS-NS follow-up of $\sim18.02$ hrs.

\textbf{OPTIMAL STRATEGY: } 
$u+g+r+i+z+y$ observations (30 s for each filter) of  NS-NS mergers with
$\Omega_{\rm{90\%}}\le100\,\rm{deg^2}$ and for which the sky position and time are favorable for LSST prompt follow up (i.e. within hours since GW trigger), followed by $g+z$ monitoring at $\delta t\ge1$ days (180 s for each filter). $u+g+r+i+z+y$ observations will start promptly and will be log-spaced in time (e.g. at 1 hr, 2hr and 4hr, and, if the field is well placed, at 8hr after merger).  We propose to continue the campaign at +1\,day and +2\,day with  $g+z$ monitoring (if at that stage we have not identified the counterpart on the first night). The prompt 6-filter monitoring will map the color evolution of the very early KN emission (and will constrain the development of new components of emission), while the later time  $g+z$ monitoring will constrain the color evolution of the red KN component and/or provide the fading signal required to identify the EM counterpart.  We propose to use deeper exposures of 180s in the  $g+z$ filters at these later epochs to track the fading of the source or to detect an intrinsically faint transient if we have not identified the EM counterpart on the first night.  During bright time, we will consider using the $r+i$ filters for the later time follow-up, which will allow us to go deeper and either increase our chances to discover the EM counterpart, or map its fading to lower fluxes. 

The $5\sigma$ magnitude limits for 30s and 180s exposures are shown in Fig. \ref{Fig:KNredblue}-\ref{Fig:KNred}. In particular, for 180s exposures we will reach $m_g^{\rm{lim}}\sim26$ mag,  $m_z^{lim}\sim24.4$ mag (ideal observing consitions, dark sky), corresponding to absolute magnitudes $M_g^{\rm{lim}}= -10.5$ and $M_z^{\rm{lim}}= -12.1$ at 200 Mpc. With this strategy, the average LSST investment of time per NS-NS merger is $\sim7.97$ hrs. The total time request for our optimal strategy to follow up NS-NS mergers with LSST is thus $47.49$ hr $\rm{yr^{-1}}$.

Our observing strategies are summarized in Table \ref{tab:strategies}.

\subsubsection{The LSST quest for the unknown: EM counterparts to Neutron Star--Black Hole mergers (NS-BH):}
As of November 2018, no GW detection of a NS-BH merger has been reported yet,  no EM counterpart to a NS-BH  merger has ever been found and no NS-BH system is known. Yet, NS-BH mergers are expected to be accompanied by KN emission not dissimilar in nature from the KN emission from NS-NS mergers, and their GW localizations are also expected to be similar \cite{2015MNRAS.446.1115M,2018LRR....21....3A}. 
The  dynamical ejecta mass produced by NS-BH mergers is theoretically predicted to be typically $\sim10$ times larger than in NS-NS mergers, leading to a more luminous KN emission peaking on average $\sim 1$ magnitude brighter than in NS-NS mergers (e.g., \cite{2015MNRAS.446.1115M}). However, the amount of lanthanide-poor ejecta is expected to be lower and, differently from NS-NS mergers, no neutron precursor is expected at early times. While some early blue emission from the disk winds is not excluded, the general expectation is that KNe associated to NS-BH mergers will be typically dominated by the NIR component \cite{2015MNRAS.446.1115M}. Since no EM counterpart to a NS-BH merger has ever been found, here we advocate for a model-agnostic observing strategy similar to NS-NS mergers of Sec. \ref{SubSubSec:NSNS}, in order to sample \emph{both} the blue and red part of the EM spectrum and either verify or challenge current theoretical predictions.

In the case of NS-BH mergers the deep sensitivity of LSST brings an additional advantage compared to all the other survey instruments. GW detectors are sensitive to NS-BH mergers at distances extending to several hundred Mpc, which implies that, on average, NS-BH mergers will be localized at larger distances than NS-NS mergers \cite{2018LRR....21....3A} (factor of a few).
The larger distances of NS-BH systems detected through their GW emission more than balance the advantage of their intrinsically more luminous KN emission. NS-BH mergers will be thus on average observed as fainter signals in the EM spectrum and will thus greatly benefit from the LSST large collecting area.

A major source of uncertainty is the intrinsic rate of yet to be observed NS-BH mergers in the local universe, which is constrained by GW observations as $R<3600\,\rm{Gpc^{-3}yr^{-1}}$ \cite{BHNSrate}, still consistent with the 90\%  confidence range of NS-NS merger rate $R\sim 300-5000\,\rm{Gpc^{-3}yr^{-1}}$ \cite{NSNSrate}. In our estimates below we follow \cite{BHNSdetectionrate} and assume an expected rate of GW detections of $\sim 10$ NS-BH mergers per year.

\textbf{MINIMAL STRATEGY:} Same as the NS-NS strategy outlined in Sec. \ref{SubSubSec:NSNS}.  The average LSST investment of time per NS-BH merger in the minimal strategy is $\sim2.17$\,hrs and $\sim2.28$\,hrs per NS-BH merger with $\Omega_{\rm{90\%}}\le20\,\rm{deg^2}$ and $20\,\rm{deg^2}<\Omega_{\rm{90\%}}\le100\,\rm{deg^2}$, respectively. We expect $\sim5\,\rm{yr^{-1}}$ NS-BH mergers to be accessible to LSST, of which $\sim3$ will have $\Omega_{\rm{90\%}}\le100\,\rm{deg^2}$. The total time request for our minimal strategy to follow up NS-BH mergers with LSST is thus $\sim 6.84$ hrs yr$^{-1}$.

\textbf{OPTIMAL STRATEGY:}  Same as the optimal NS-NS strategy outlined in Sec. \ref{SubSubSec:NSNS}. The LSST investment of time per NS-BH merger with  $\Omega_{\rm{90\%}}<100\,\rm{deg^2}$ is 7.97 hrs. The expected rate of events satisfying these criteria is  $\sim$3\,$\rm{yr^{-1}}$, leading to a yearly time request of $\sim 23.89$ hrs to follow up NS-BH mergers with LSST.

Our observing strategies are summarized in Table \ref{tab:strategies}.

\subsubsection{The LSST quest for the unknown: EM counterparts to Black-Hole Black-Hole mergers (BH-BH)}
Theoretical speculations on EM counterparts to BH-BH mergers recently experienced a surge of interest because of the possible association of a burst of $\gamma$-rays detected by the Fermi satellite with the BH-BH merger event GW150914 \cite{Connaughton16}.
BH-BH mergers are routinely detected by LIGO/Virgo through their GW emission, but to date an unambiguous association with an EM counterpart is still missing. Theoretical models of EM counterparts from BH-BH mergers are highly speculative and span a wide range of possible morphologies \cite{Perna,Loeb,Stone,deMink,McKernan}. On the observational side, the large localization regions of two-interferometer detections have prevented searches for optical BH-BH emission from being complete in color, space, and time. As such, the existence and properties of EM transient emission from BH-BH mergers is still a completely open question in astrophysics.  Given the current large uncertainty of possible EM counterparts, we design a model-agnostic LSST observational strategy of nearby, well localized BH-BH mergers with the goal to conduct a systematic search for EM emission from BH-BH mergers down to a level of sensitivity that is simply not possible with any other ground-based survey instrument.

Following \cite{GW170104} we adopt a BH-BH merger rate $R\sim10-200\,\rm{Gpc^{-3}yr^{-1}}$ and design a ``minimal'' and ``optimal'' LSST observing strategy based on the results from BH-BH merger simulations as seen through the eyes of GW detectors \cite{PankowSkyLocInPrep}. These simulations allow us to have a realistic prediction of the typical size of GW localization regions as a function of the BH-BH merger distance and detectors duty cycle (which is conservatively assumed to be uncorrelated among detectors). 

\textbf{MINIMAL STRATEGY:} LSST follow up of promptly accessible (i.e. within hours of GW detection) BH-BH mergers at $d_L\le500$ Mpc  with $\Omega_{90\%}\le50\,\rm{deg^2}$. The expected rate of BH-BH mergers satisfying the criteria above and accessible to LSST is $\le$ a few $\rm{yr^{-1}}$, based on \cite{PankowSkyLocInPrep}. On average, we expect to be able to cover the GW localization region with $\sim$10 LSST pointings. We advocate for follow up in two LSST filters that span the widest possible range of the electromagnetic spectrum, while maximizing the depth of our search for EM counterparts.  We propose deep  $g+i$ observations during dark time and $r+i$ observations during bright time ($180$ s exposure for each filter) to avoid throughput losses of the $u$ and $y$ filter. 

We propose deep $g+i$ (or $r+i$ during bright time) observations (180 s exposure for each filter) at 1 hr, 3 days and 15 days after the merger. The average investment of LSST time per BH-BH merger is $3.28$\,hr (total of $\sim6.56$ hr $\rm{yr^{-1}}$). For a 180s exposure observation anticipate reaching a $5\sigma$ magnitude limit  $m_g^{\rm{lim}}\sim26$  mag  $m_i^{\rm{lim}}\sim25$ mag (under ideal conditions of dark sky and zenith pointing), corresponding to absolute magnitudes $M_g^{\rm{lim}}= -12.5$ and $M_i^{\rm{lim}}= -13.5$ at 500 Mpc.

\textbf{OPTIMAL STRATEGY:} Same as the the minimal strategy outlined above, but with the addition of another epoch of deep $g+i$  observations  (or $r+i$ during bright time) during the first night. This strategy will allow us to map the very short time-scales of variability of potential EM transients associated with BH-BH mergers, as well as the longer time scales of evolution of $\sim$\,weeks.  The average investment of LSST time per BH-BH merger is $4.38$\,hr (total of $\sim8.76$ hr $\rm{yr^{-1}}$).

Equipped with ToO capabilities, LSST will probe the existence and properties of transients from BH-BH mergers with unparalleled sensitivity among ground-based surveys, thus opening up a completely new window of investigation on our Universe. As a comparison, at the time of writing the optical follow-up campaign with the highest GW sky map probability coverage of BH-BH mergers is described in \cite{Doctor}. This analysis searched for fast-fading emission with $i$-band observations of the GW178014 localization region at $\delta t\leq12$ days down to an absolute mag $M_i^{lim}\sim-15.5$. With the BH-BH follow-up campaign we  have proposed here, under ideal observing conditions LSST will extend the discovery space $\sim3$ magnitudes deeper, probing fast and slow time scales of evolution of EM counterparts to BH-BH mergers in two bands (hence providing color information).
The key advantage of the ideal strategy above, compared to the minimal strategy, is the capability to sample the very short time scales of evolution of the transients.

Our observing strategies are summarized in Table \ref{tab:strategies}.

\subsubsection{The LSST quest for the unknown: unidentified GW sources}
\label{SubSubSec:GWunknown}
This class of GW triggers include sources found through the GW burst pipeline, which are not necessarily of compact-object merger origin and might include very nearby supernova explosions and things we may not even have thought of. 
We propose LSST follow-up of $\sim2-3$ unidentified GW sources per year with localization $\Omega_{\rm{90\%}}\le 100$\,deg$^2$. We expect to be able to cover the localization region with $\sim20$ LSST repointings.  We propose $g+z$ 30s-exposure observations during the first night, at 3 days and 15 days to sample the EM spectrum with deep sensitivity ($r+z$ will be used during bright time). For GW sources ideally placed in the sky, two $g+z$ epochs will be acquired during the first night. With this strategy, we will be able to constrain the presence of EM counterparts to unidentified GW sources across the spectrum, both on short (i.e. intra-night) and longer time-scales of weeks.  The average investment of time per GW trigger is $1.52$\,hr, which is a yearly investment of $4.56$\,hr of LSST. This is a small investment of $\sim 0.1\%$ of LSST time, which hold promises for high discovery potential and significant scientific impact.

Our observing strategies are summarized in Table \ref{tab:strategies}.

\subsection{Footprint -- pointings, regions and/or constraints}
\begin{footnotesize}{\it Describe the specific pointings or general region (RA/Dec, Galactic longitude/latitude or 
Ecliptic longitude/latitude) for the observations. Please describe any additional requirements, especially if there
are no specific constraints on the pointings (e.g. stellar density, galactic dust extinction).}
\end{footnotesize}

For each GW event that we follow up we will nominally tile the 90\% confidence region of the GW sky map with filters of interest and the times described in \S 3.1 and summarized in Table \ref{tab:strategies}. This strategy will result in 10\% of sources being missed.  We may tile more than 90\% of the localization region if the resources are available, but will maintain the same pointings GW event to GW event. The pointings can be taken in any order on the sky, since few pointings are needed to tile the area and each exposure is short.  Ideally, each tiling would happen all at once and not spread over the entirety of a night.

\subsection{Image quality}
\begin{footnotesize}{\it Constraints on the image quality (seeing).}\end{footnotesize}

No constraints on image quality.
\subsection{Individual image depth and/or sky brightness}
\begin{footnotesize}{\it Constraints on the sky brightness in each image and/or individual image depth for point sources.
Please differentiate between motivation for a desired sky brightness or individual image depth (as 
calculated for point sources). Please provide sky brightness or image depth constraints per filter.}
\end{footnotesize}

For each strategy we outlined the ideal exposure times and filters, depending on the observing conditions, type of event and time since GW trigger as described in \S 3.1 and summarized in Table \ref{tab:strategies}. We emphasize that the primary driver of our scientific case is timeliness. 

\subsection{Co-added image depth and/or total number of visits}
\begin{footnotesize}{\it  Constraints on the total co-added depth and/or total number of visits.
Please differentiate between motivations for a given co-added depth and total number of visits. 
Please provide desired co-added depth and/or total number of visits per filter, if relevant.}
\end{footnotesize}

No constraints on co-added image depth or total number of visits.
\subsection{Number of visits within a night}
\begin{footnotesize}{\it Constraints on the number of exposures (or visits) in a night, especially if considering sequences of visits.  }
\end{footnotesize}

For GW sources we choose to follow up, we will attempt to take exposures of the 90\% localization region that are logarithmically spaced in time.  Ideally, this means two or more visits to the region on the first night after the GW trigger and at least one visit the following night.  Later observations, if needed to identify the counterpart, would only tile the localization region once in a night and could be flexibly timed to accommodate other programs. 

\subsection{Distribution of visits over time}
\begin{footnotesize}{\it Constraints on the timing of visits --- within a night, between nights, between seasons or
between years (which could be relevant for rolling cadence choices in the WideFastDeep. 
Please describe optimum visit timing as well as acceptable limits on visit timing, and options in
case of missed visits (due to weather, etc.). If this timing should include particular sequences
of filters, please describe.}
\end{footnotesize}

For each GW source we follow up, we plan to acquire observations that are logarithmically spaced in time. Details are provided in \S 3.1 and Table \ref{tab:strategies}.  In particular, for most of our targets we propose multiple visits during the first night of observation, which are crucial to identify the EM counterpart and constrain its early time behavior, followed by observations in the following days. We will discontinue the LSST follow up in the case of unambiguous identification of the EM counterpart by LSST or other facilities. At this point, telescopes with smaller field of view will take over.

\subsection{Filter choice}
\begin{footnotesize}
{\it Please describe any filter constraints not included above.}
\end{footnotesize}

Described in \S 3.1 and Table \ref{tab:strategies}.

\subsection{Exposure constraints}
\begin{footnotesize}
{\it Describe any constraints on the minimum or maximum exposure time per visit required (or alternatively, saturation limits).
Please comment on any constraints on the number of exposures in a visit.}
\end{footnotesize}

Optical counterparts to NS-NS and NS-BH systems are expected to quickly brighten over a few hours after the merger and then fade over a typical time-scale of a couple of weeks.  We propose matching these light curves by performing 30s exposures during the first night, and then switching to 180s exposures afterwards.  Since BH-BH merger EM emission is poorly constrained by current observations and theory, we aim to maximize depth for BH-BH follow-up.  As such, all BH-BH follow-up exposures will be 180s. For unidentified GW sources, we adopt 30s exposures. Details for each strategy are provided in \S 3.1 and Table \ref{tab:strategies}.

\subsection{Other constraints}
\begin{footnotesize}
{\it Any other constraints.}
\end{footnotesize}

Ideally, LSST will publicly release the ToO follow-up strategy on each event in advance to maximize the opportunity for coordination with other observatories, including optical photometric/spectroscopic facilities on the ground and from space, as well as observations outside the optical band.

Optical-NIR ground-based observatories of interest include:  (i) the  Southern Astrophysical Research (SOAR)  and Gemini-South telescopes, which share with LSST the Cerro Pachon area (hence ideally positioned to help identify the EM counterpart among potential candidates, and take over the LSST observing campaign once the EM counterpart has been identified); (ii) ESO-NTT telescope in La Silla equipped with the SOXS (Son Of X-Shooter) spectrograph \cite{SOXS}. SOXS is a high-efficiency spectrograph with wide spectral range with focus on spectroscopic follow-up of transient sources (expected to start of operations in 2021).
Future space-based observatories of interest include the Time-Domain Spectroscopic Observatory (TSO), an optical-IR imaging and spectroscopic telescope that would provide time-critical spectroscopy information on EM candidates for identification and characterization.

\subsection{Estimated time requirement}
\begin{footnotesize}
{\it Approximate total time requested for these observations, using the guidelines available at \url{https://github.com/lsst-pst/survey_strategy_wp}.}
\end{footnotesize}

Our total time request for the MINIMAL strategy is $\sim36$\,hr yr$^{-1}$ ($\sim1\%$ of LSST time). For the OPTIMAL STRATEGY our time request is  $\sim85$\,hr yr$^{-1}$ ($\sim2\%$ of LSST time). We provide details of our time request in \S 3.1, with a summary in Table \ref{tab:strategies}.

\begin{sidewaystable}
    \centering
    \begin{tabular}{r|l|l|l|}
        \toprule
         & \textbf{MINIMAL Strategy NS-NS} & \textbf{OPTIMAL Strategy NS-NS} \hspace{.3in}\\
        \midrule
        \textbf{Sequence} & $ugrizy$ (30s) at 1h, 2h, 4h, (8h), 1d, (2d) & $ugrizy$ (30s) at 1h, 2h, 4h, (8h),  $gz^{*}$ (180s) at 1d,\\ 
        & for $\Omega_{\rm{90\%}}\le20$deg$^2$ &  (2d) for $\Omega_{\rm{90\%}}\le100$deg$^2$\\
         & $gz$ (30s) at 1h, 2h, 4h, (8h), 1d, (2d)   & \\
         & for $20$deg$^{2}<\Omega_{\rm{90\%}}\le100$deg$^2$ & \\
        \textbf{Average Time per Event} & 2.17 hrs for $\Omega_{\rm{90\%}}\le20$deg$^2$ &  7.97 hrs\\
        & 2.28 hrs for $20$deg$^{2}<\Omega_{\rm{90\%}}\le100$deg$^2$& \\
        \textbf{Events per year} & (1-2) with $\Omega_{\rm{90\%}}\le20$deg$^2$ &  (1-10)\\
        &(1-10) with $20$deg$^{2}<\Omega_{\rm{90\%}}\le100$deg$^2$  & \\
        \textbf{Total Time per LSST year} & 18.02\,hrs & 47.79\,hrs\\
          \toprule
       & \textbf{MINIMAL Strategy NS-BH} & \textbf{OPTIMAL Strategy NS-BH} \hspace{.3in}\\
               \toprule
\textbf{Sequence} & Same as for NS-NS & Same as for NS-NS\\
        \textbf{Average Time per Event} & Same as for NS-NS & Same as for NS-NS\\
 \textbf{Events per year}&$\sim3$ & $\sim3$\\
\textbf{Total Time per LSST year}&  6.84\,hrs & 23.89\,hrs\\
          \toprule
      & \textbf{MINIMAL Strategy BH-BH} & \textbf{OPTIMAL Strategy BH-BH} \hspace{.3in}\\
               \toprule
\textbf{Sequence} & $gi^{*}$ (180 s ) at 1h, 3d, 15d for $\Omega_{\rm{90\%}}\le50$deg$^2$ & $gi^{*}$ (180 s ) at 1h, 4h, 3d, 15d for $\Omega_{\rm{90\%}}\le50$deg$^2$\\
\textbf{Average Time per Event} & 3.28\,hr & 4.38\,hr\\
\textbf{Events per year}& $\sim2$ & $\sim2$\\
\textbf{Total Time per LSST year}& 6.56\,hr & 8.76\,hr\\
         \toprule
      & \textbf{MINIMAL Strategy Unidentified GW} & \textbf{OPTIMAL Strategy Unidentified GW} \hspace{.3in}\\
               \toprule
                    \textbf{Sequence} &Same as optimal& $gz^{*}$ (30s) at 1h, 3d, 15d\\
     \textbf{Average Time per Event}&Same as optimal&1.52 hr\\
     \textbf{Events per year} &Same as optimal& (2-3)\\
  \textbf{Total Time per LSST year}&  Same as optimal & 4.56 hr\\   
          \toprule
        \toprule
     & \textbf{MINIMAL Strategy - Total Request} & \textbf{OPTIMAL Strategy  - Total Request} \hspace{.3in}\\
      & 35.98 hr ($\sim$1.0\% LSST time) & 85.00 hr ($\sim$2.3\% LSST time)\\
        \bottomrule
    \end{tabular}
    \caption{{\bf Summary of Strategies and Time request.} ($^*$) we will use $r$-band instead of $g$-band during bright time. }
        \label{tab:strategies}
\end{sidewaystable}

\begin{table}[ht]
    \centering
    \begin{tabular}{l|l|l|l}
        \toprule
        Properties & Importance \hspace{.3in} \\
        \midrule
        Image quality &   3  \\
        Sky brightness &  3\\
        Individual image depth &  3 \\
        Co-added image depth &  3 \\
        Number of exposures in a visit   & 3  \\
        Number of visits (in a night)  &  2 \\ 
        Total number of visits &  2 \\
        Time between visits (in a night) &  1\\
        Time between visits (between nights)  & 1  \\
        Long-term gaps between visits & 3\\
        Filters used in visits & 1 \\
        Timeliness of response to trigger & 1\\
        \bottomrule
    \end{tabular}
    \caption{{\bf Constraint Rankings:} Summary of the relative importance of various survey strategy constraints. Please rank the importance of each of these considerations, from 1=very important, 2=somewhat important, 3=not important. If a given constraint depends on other parameters in the table, but these other parameters are not important in themselves, please only mark the final constraint as important. For example, individual image depth depends on image quality, sky brightness, and number of exposures in a visit; if your science depends on the individual image depth but not directly on the other parameters, individual image depth would be `1' and the other parameters could be marked as `3', giving us the most flexibility when determining the composition of a visit, for example.}
        \label{tab:obs_constraints}
\end{table}

\subsection{Technical trades}
\begin{footnotesize}
{\it To aid in attempts to combine this proposed survey modification with others, please address the following questions:
\begin{enumerate}
    \item What is the effect of a trade-off between your requested survey footprint (area) and requested co-added depth or number of visits?
    \item If not requesting a specific timing of visits, what is the effect of a trade-off between the uniformity of observations and the frequency of observations in time? e.g. a `rolling cadence' increases the frequency of visits during a short time period at the cost of fewer visits the rest of the time, making the overall sampling less uniform.
    \item What is the effect of a trade-off on the exposure time and number of visits (e.g. increasing the individual image depth but decreasing the overall number of visits)?
    \item What is the effect of a trade-off between uniformity in number of visits and co-added depth? Is there any benefit to real-time exposure time optimization to obtain nearly constant single-visit limiting depth?
    \item Are there any other potential trade-offs to consider when attempting to balance this proposal with others which may have similar but slightly different requests?
\end{enumerate}}
\end{footnotesize}

Any major modification of the footprint (area), exposure times, filters and times of observations that we propose here would have a highly disruptive impact on our capability to reach our scientific objectives. The impact of this ToO program on other programs can be mitigated if observations acquired as ToOs can be used as part of other LSST surveys.

\section{Performance Evaluation}
\begin{footnotesize}
{\it Please describe how to evaluate the performance of a given survey in achieving your desired
science goals, ideally as a heuristic tied directly to the observing strategy (e.g. number of visits obtained
within a window of time with a specified set of filters) with a clear link to the resulting effect on science.
More complex metrics which more directly evaluate science output (e.g. number of eclipsing binaries successfully
identified as a result of a given survey) are also encouraged, preferably as a secondary metric.
If possible, provide threshold values for these metrics at which point your proposed science would be unsuccessful 
and where it reaches an ideal goal, or explain why this is not possible to quantify. While not necessary, 
if you have already transformed this into a MAF metric, please add a link to the code (or a PR to 
\href{https://github.com/lsst-nonproject/sims_maf_contrib}{sims\_maf\_contrib}) in addition to the text description. }
\end{footnotesize}

The LSST main survey (WFD) has an overall efficiency of KN detection of only a few \%, with most light-curves of recovered KNe consisting of sparse observations that preclude accurate constrains on the physical parameters of interest \cite{Cowperthwaite18}.
Instead, with the LSST ToO optimal strategy outlined above 
we expect an EM counterpart discovery in the vast majority of NS-NS mergers.  
With the LSST minimal ToO strategy we anticipate a lower level of success
(e.g.\, less timely EM candidate identification, which might prevent subsequent characterization of the source with smaller FOV facilities, or limited  information on the early time properties of the EM counterpart, which will preclude the identification of additional components of emission). Based on these considerations, we define a heuristic quantifier of the success of the ToO implementation for NS-NS merger follow-up as:
\begin{equation}
S_\mathrm{NS-NS} = \frac{(1+2 f_\mathrm{early})N_\mathrm{det}}{3 N_\mathrm{NS-NS}}
\end{equation}
where $N_\mathrm{NS-NS}$ is the number of NS-NS mergers detected by GW interferometers that satisfy the ToO activation criteria, $N_\mathrm{det}$ is the number of associated KN detections in LSST ToOs, and $f_\mathrm{early}$ is the fraction of these detections that lead to an identification within 1 day. With this definition (which considers an early detection three times as valuable as a later one), the success metric can take values between 0 (absolute failure, no detections at all) to 1 (complete success, all counterparts detected within 1 day from GW trigger).

We provide below a rough estimate of our expected performance with the optimal and minimal strategy. The number $N_\mathrm{det}$ of KN detections from NS-NS mergers is proportional to the number of ToOs following up that type of events, and to the detection efficiency $f_\mathrm{det}$, i.e. 
\begin{equation}
N_\mathrm{det}\sim \frac{T_\mathrm{ToO}}{<T_\mathrm{single}>}f_\mathrm{NS}f_\mathrm{det} 
\end{equation}
where $T_\mathrm{ToO}$ is the total time allocated to ToOs, $<T_\mathrm{single}>$ is the average time invested on ToOs per event, and $f_\mathrm{NS}$ is the fraction of ToOs time dedicated to NS-NS events in the two strategies. Within the first year of operations, we expect $N_\mathrm{NS-NS}\sim 1-10$ (Table~\ref{tab:strategies}). 

For the optimal strategy, $<T_\mathrm{single}>\sim 8\,\mathrm{hr}$, $T_\mathrm{ToO}=80\,\mathrm{hr}$ and $f_\mathrm{NS}=0.59$. From Figure 11 of \cite{Cowperthwaite18}, we can see that the $180$ s exposures of the optimal strategy lead to $f_\mathrm{det}\sim 1$. Assuming  $f_\mathrm{early}\sim 1$, we have $S_\mathrm{NS-NS}\sim 0.6$ for the optimal strategy.

For the minimal strategy, $<T_\mathrm{single}>\sim 2.3\,\mathrm{hr}$, $T_\mathrm{ToO}=40\,\mathrm{hr}$ and $f_\mathrm{NS}=0.57$. From Figure 11 of \cite{Cowperthwaite18}, we see that keeping the exposures $30$ s long leads to $f_\mathrm{det}\sim 0.5$ at 200 Mpc. Assuming  $f_\mathrm{early}\sim 0.5$, we obtain $S_\mathrm{NS-NS}\sim 0.33$ for the minimal strategy. \textit{We thus set $S_\mathrm{NS-NS}\sim 0.33$ as the minimum requirement on $S_\mathrm{NS-NS}$ for our science goal to be considered successful.} After each year of LSST operation, this number can be computed using the real number of detections and efficiencies from that year, and it can be used to quantify the success of ToOs in the follow-up of NS-NS mergers. 

For NS-BH mergers, BH-BH mergers and GW events from unidentified sources for which an optical-NIR EM counterpart has never been observed, defining the rate of success of our strategy in a similar, semi-quantitative way is not straightforward, as in this case LSST is literally exploring the unknown.
However, we emphasize that those EM counterparts constitute a large portion of the discovery space that is made available for LSST exploration by our ToO strategies.

\textbf{Finally, we strongly advocate for a re-evaluation of the LSST ToO triggering criteria and observing strategies that we propose here after the completion of LIGO/Virgo O3 in 2019, and on a yearly base after the start of LSST operations.}
 
\section{Special Data Processing}
\begin{footnotesize}
{\it Describe any data processing requirements beyond the standard LSST Data Management pipelines and how these will be achieved.}
\end{footnotesize}
\begin{itemize}
\item The success of our program largely relies on the capability to rapidly identify EM counterparts of GW triggers in real time. For this reason, in addition to standard Level 1 data products, we request that images of our search region be promptly available (ideally as soon as acquired), to allow the astronomical community to run optimized search algorithms  (including real time stacking of overlapping images) and distribute alerts about potential EM candidates to other observatories. 
\item Stacking of images of our search region acquired during the same night made and  promptly available to the astronomical community will also be greatly beneficial to our search for the faintest EM counterparts.
\item Finally, \textbf{we emphasize that the possibility to ``re-integrate'' images acquired as part of the ToO program within other LSST surveys will greatly reduce the impact of ToOs on other scientific cases}. 
\end{itemize}

* This work developed partly within the TVS Science Collaboration and the author acknowledge the great support of TVS in the preparation of this paper. 

** The authors acknowledge support from the Flatiron Institute and Heising-Simons Foundation for the development of this paper.

*** To our knowledge, two other LSST White Papers (WP)  include a discussion of KNe or GW follow up (the WP from LSST-DESC and the WP by Andreoni et al.).

\section{References}

\bibliographystyle{hunsrt}
\begingroup
\renewcommand{\section}[2]{}%
\bibliography{refs}
\endgroup

\end{document}